\documentclass[12pt]{article}
\usepackage{amsfonts}
\usepackage{amssymb}
\usepackage{fancyhdr}
\usepackage{slashed}
\usepackage{graphicx}
\usepackage{subfigure}
\usepackage{color}

\usepackage{hyperref}
\def\hybrid{\topmargin -20pt    \oddsidemargin 0pt
        \headheight 0pt \headsep 0pt
        \textwidth 6.25in       % A4 paper
        \textheight 9.25in       % A4 paper
        \marginparwidth .875in
        \parskip 5pt plus 1pt   \jot = 1.5ex}

\hybrid

\def\baselinestretch{1.2}

\catcode`\@=11

\def\marginnote#1{}
\newcount\hour
\newcount\minute
\newtoks\amorpm
\hour=\time\divide\hour by60
\minute=\time{\multiply\hour by60 \global\advance\minute by-\hour}
\edef\standardtime{{\ifnum\hour<12 \global\amorpm={am}%
        \else\global\amorpm={pm}\advance\hour by-12 \fi
        \ifnum\hour=0 \hour=12 \fi
        \number\hour:\ifnum\minute<10 0\fi\number\minute\the\amorpm}}
\edef\militarytime{\number\hour:\ifnum\minute<10 0\fi\number\minute}
\def\draftlabel#1{{\@bsphack\if@filesw {\let\thepage\relax
   \xdef\@gtempa{\write\@auxout{\string
      \newlabel{#1}{{\@currentlabel}{\thepage}}}}}\@gtempa
   \if@nobreak \ifvmode\nobreak\fi\fi\fi\@esphack}
        \gdef\@eqnlabel{#1}}
\def\@eqnlabel{}
\def\@vacuum{}
\def\draftmarginnote#1{\marginpar{\raggedright\scriptsize\tt#1}}

\def\draft{\oddsidemargin -.5truein
        \def\@oddfoot{\sl preliminary draft \hfil
        \rm\thepage\hfil\sl\today\quad\militarytime}
        \let\@evenfoot\@oddfoot \overfullrule 3pt
        \let\label=\draftlabel
        \let\marginnote=\draftmarginnote
   \def\@eqnnum{(\theequation)\rlap{\kern\marginparsep\tt\@eqnlabel}%
\global\let\@eqnlabel\@vacuum}  }

\def\preprint{\twocolumn\sloppy\flushbottom\parindent 2em
        \leftmargini 2em\leftmarginv .5em\leftmarginvi .5em
        \oddsidemargin -.5in    \evensidemargin -.5in
        \columnsep .4in \footheight 0pt
        \textwidth 10.in        \topmargin  -.4in
        \headheight 12pt \topskip .4in
        \textheight 6.9in \footskip 0pt
        \def\@oddhead{\thepage\hfil\addtocounter{page}{1}\thepage}
        \let\@evenhead\@oddhead \def\@oddfoot{} \def\@evenfoot{} }

\def\numberbysection{\@addtoreset{equation}{section}
        \def\theequation{\thesection.\arabic{equation}}}

\def\underline#1{\relax\ifmmode\@@underline#1\else
        $\@@underline{\hbox{#1}}$\relax\fi}

\def\titlepage{\@restonecolfalse\if@twocolumn\@restonecoltrue\onecolumn
     \else \newpage \fi \thispagestyle{empty}\c@page\z@
        \def\thefootnote{\fnsymbol{footnote}} }

\def\endtitlepage{\if@restonecol\twocolumn \else \newpage \fi
        \def\thefootnote{\arabic{footnote}}
        \setcounter{footnote}{0}}  %\c@footnote\z@ }

\catcode`@=12
\relax

\def\figcap{\section*{Figure Captions\markboth
        {FIGURECAPTIONS}{FIGURECAPTIONS}}\list
        {Figure \arabic{enumi}:\hfill}{\settowidth\labelwidth{Figure
999:}
        \leftmargin\labelwidth
        \advance\leftmargin\labelsep\usecounter{enumi}}}
 \relax
\def\tablecap{\section*{Table Captions\markboth
        {TABLECAPTIONS}{TABLECAPTIONS}}\list
        {Table \arabic{enumi}:\hfill}{\settowidth\labelwidth{Table
999:}
        \leftmargin\labelwidth
        \advance\leftmargin\labelsep\usecounter{enumi}}}
 \relax
\def\reflist{\section*{References\markboth
        {REFLIST}{REFLIST}}\list
        {[\arabic{enumi}]\hfill}{\settowidth\labelwidth{[999]}
        \leftmargin\labelwidth
        \advance\leftmargin\labelsep\usecounter{enumi}}}
 \relax
\makeatletter
\newcounter{pubctr}
\def\publist{\@ifnextchar[{\@publist}{\@@publist}}
\def\@publist[#1]{\list
        {[\arabic{pubctr}]\hfill}{\settowidth\labelwidth{[999]}
        \leftmargin\labelwidth
        \advance\leftmargin\labelsep
        \@nmbrlisttrue\def\@listctr{pubctr}
        \setcounter{pubctr}{#1}\addtocounter{pubctr}{-1}}}
\def\@@publist{\list
        {[\arabic{pubctr}]\hfill}{\settowidth\labelwidth{[999]}
        \leftmargin\labelwidth
        \advance\leftmargin\labelsep
        \@nmbrlisttrue\def\@listctr{pubctr}}}
 \relax
\makeatother
\newskip\humongous \humongous=0pt plus 1000pt minus 1000pt

\newif\ifdtup

\relax

\def\be{\begin{equation}}
\def\ee{\end{equation}}
\def\ba{\begin{eqnarray}}
\def\ea{\end{eqnarray}}

\def\del{\partial}

\def\a{\alpha}

\def\b{\beta}

\def\g{\gamma}

\def\d{\delta}

\def\e{\epsilon}

\def\th{\theta}

\def\l{\lambda}

\def\s{\sigma}

 \def\cE{{\cal E}}

\newcommand{\vev}[1]{{\left< {#1} \right>}}

\newcommand{\prt}[1]{{\left( {#1} \right)}}

\def\no{\noindent}

\def\IR{\relax{\rm I\kern-.18em R}}

\def\IR{\relax{\rm I\kern-.18em R}}
\def\IL{\relax{\rm I\kern-.18em L}}

\def\inv{^{\raise.15ex\hbox{${\scriptscriptstyle -}$}\kern-.05em 1}}

\def\cE{{\cal E}}

\def\bea{\begin{eqnarray}}
\def\eea{\end{eqnarray}}
\newcommand{\eq}[1]{(\ref{#1})}
\def\nn{\nonumber}
\def\del{\partial}
\newcommand{\la}[1]{\label{#1}}

\def\a{\alpha}      
\def\b{\beta}       
\def\g{\gamma}    
\def\d{\delta}    
\def\e{\epsilon}

\def\l{\lambda} 
 
\def\o{\omega}

\def\s{\sigma}  
\def\t{\tau}
\def\th{\theta}

\definecolor{markcolor2}{rgb}{1,0,0}

\definecolor{markcolor3}{rgb}{0,1,0}

\begin{document}

\renewcommand{\theequation}{\thesection.\arabic{equation}}
\csname @addtoreset\endcsname{equation}{section}

\newcommand{\beq}{\begin{equation}}
\newcommand{\eeq}[1]{\label{#1}\end{equation}}
\newcommand{\ber}{\begin{eqnarray}}
\newcommand{\eer}[1]{\label{#1}\end{eqnarray}}
\newcommand{\eqn}[1]{(\ref{#1})}

\begin{titlepage}

\begin{center}

%\hfill CALT-xx-yyyy\\
%\vskip -.1 cm
%\hfill hep--th/yymmnnn\\

~
\vskip 1 cm

{\Large
\bf On the holographic width of flux tubes}

\vskip 0.5in

{\bf Dimitrios Giataganas$^{1,2}$\phantom{x}and\phantom{x}Nikos Irges}$^{3}$
\vskip 0.1in
{\em
{}$^1$Department of Nuclear and Particle Physics,\\
Faculty of Physics, University of Athens,\\
Athens 15784, Greece
\vskip .15in
{}$^2$ Rudolf Peierls Centre for Theoretical Physics,\\
University of Oxford, 1 Keble Road,\\
Oxford OX1 3NP, United Kingdom\\
\vskip .15in
3. Department of Physics, National Technical University of Athens\\
  Zografou Campus, GR-15780 Athens, Greece
}
\vskip .1in
{\small \sffamily
dgiataganas@phys.uoa.gr, irges@mail.ntua.gr
\\
}
\vskip .2in
\end{center}

\vskip .4in

\centerline{\bf Abstract}

We investigate the width of the flux tube between heavy static quark charges. Using the
gauge/gravity duality, we find the properties of the minimal connected surface   related to the width 
of the bound state. We show that in the confining phase, the logarithmic broadening predicted by the effective 
string description and observed in lattice simulations is a generic property of all confining backgrounds. 
We also study the transverse fluctuations of the string connecting two static quarks in
curved backgrounds. Our formalism is applied to $AdS$ space where we compute the expectation value 
of the square of transverse deviations of the string, a quantity related to the width.

\no
\end{titlepage}
\vfill
\eject

%\end{center}

\noindent

%\vskip .4in
%\noindent
%August 2002\\
%\end{titlepage}
%\vfill
%\eject

\def\baselinestretch{1.2}
\baselineskip 19 pt
\noindent

%%%%%%%%%%%%%%%

\setcounter{equation}{0}

\tableofcontents
%%%%%%%%%%%%%%%%%%
\section{Introduction}
%%%%%%%%%%%%%%%%%%

The confinement property of the strong force is of fundamental interest and while a lot of effort has been devoted to its study,
our understanding of the non-perturbative dynamics involved is far from being complete.
A way to study confinement is to use as probe a heavy bound state of a quark and an anti-quark. The fundamental quantity that one can compute using the pair is the expectation value of the rectangular Wilson loop.
The shape of the loop is related to the fact that the probes are assumed to be very massive and that through time they stay at fixed positions.
The two long legs of the Wilson loop correspond to the time direction, while its other two sides are along the alignment of the pair
representing the distance between the color charged particles. In a confining theory the Wilson loop gives the known area law,
which is equivalent to a linear static potential in the inter-quark distance. The expectation value of the Wilson loop is therefore considered to be a criterion for confinement.  At low temperatures, below the deconfinement transition temperature, string models predict a decrease of the potential and in particular the string tension decreases as temperature increases  \cite{Pisarski:1982cn, deForcrand:1984cz, Gao:1989kg}, fact that has been also confirmed recently using the gauge/gravity duality \cite{Giataganas:2014mla}. 

There exists another quantity of almost equal importance that has not attracted the same attention as the potential of confining theories.
It is the chromo-electric flux tube formed between the quark pair. Due to confinement, the chromo-electric field energy density is believed to be
constrained in a tube-like region. In confining theories the tube has some interesting universal properties. Using a low-energy effective string
description it has been shown that the flux tube width grows logarithmically with the size of the bound state, when measured at the middle
of the pair  \cite{Luscher:1980iy}. This effect, sometimes called roughening, together with the appearance of the L\"{u}scher term signals
the quantum string-like behavior of the tube joining the pair. The rough phase is described by strong fluctuations of the collective coordinates of the position of the string and the mean squared width of the flux tube diverges logarithmically when the inter-quark distance goes to infinity. The string like features of the confining flux tube have been studied in several gauge models, summarized in recent reviews, including \cite{Teper:2009uf,Lucini:2012gg,Panero:2012qx}.
The broadening is believed to occur due to the transverse fluctuations of gluonic strings that increase in magnitude as the quarks are separated. At zero temperature, the logarithmically divergent string width has been verified by several lattice computations with increasing accuracy \cite{Caselle:1995fh,Gliozzi:2010zv,Caselle:2012rp}.
These results have been extended by the study of the thickness of the tube in higher representations of the gauge groups. It has been found that the width
of the flux tube of the bound state grows logarithmically due to the excitations of the underlying k-strings \cite{Giudice:2006hw}.
Moreover, the width of the baryonic flux-tube junction from effective string theory has also been found to grow logarithmically \cite{Pfeuffer:2008mz}. It is interesting to point
out that logarithmic like behavior has been observed in several other similar models, in statistical physics and condensed matter physics, therefore hinting to universal properties.
A more recent result is that the roughening undergoes a qualitative change at finite temperature slightly before the deconfinement phase transition, as
the broadening apparently changes from the zero temperature logarithmic law to a linear law \cite{Gliozzi:2010zv,Gliozzi:2010zt,Gliozzi:2010jh} as the phase transition is approached. All these properties of confinement can be studied holographically confirming the lattice theory results and with a hope that new unknown properties will be revealed.

There are more than one equivalent ways to measure the effective width of the flux tube.
In lattice studies, the computation is done by considering two Wilson loops. One Wilson loop $W\prt{C}$
corresponds to the actual heavy $Q {\bar Q}$ bound state that generates the flux tube, with one large time direction and one smaller
spatial direction fixed to be the size of the bound state. The other, small probe Wilson loop $P\prt{c}$ is placed on a plane above the $Q {\bar Q}$ Wilson loop plane in the center and measures the flux passing through it \cite{Luscher:1980iy}.
The chromoelectric field energy density $\cE\prt{x}$ is defined as
\be\la{wdef1}
\cE\prt{x}=\frac{\vev{W\prt{C} P\prt{c}}-\vev{W\prt{C}}\vev{P\prt{c}}}{\vev{W\prt{C}}}~
\ee
and recall that in holography it is specified by the minimal surface connecting the two Wilson loops.
The mean square width of the flux tube is then defined as
\be\la{w2def}
w^2=\frac{\int d x_\perp x_\perp^2 \cE\prt{x}}{\int d x_\perp \cE\prt{x}}\, ,
\ee
where $x_\perp$ are the plane dimensions transverse to the direction of the heavy pair.

Another way to calculate the width of the flux tube is through the time-dependent transverse fluctuations of the string connecting 
the static quarks with appropriate Dirichlet boundary conditions for the spatial direction. 
In that case the expectation value of the square transverse deviations of the string has been studied extensively in \cite{Gliozzi:2010zv,Gliozzi:2010zt} and is given by
\bea
w^2(x) &=& \langle \chi(x,t)^2 \rangle \equiv \lim_{\e\to 0}\langle \chi (x,t) \chi (x'=x+\e, t'=t+\e) \rangle\nonumber\\
&=&  \lim_{\e\to 0}G(x,t;x'=x+\e,t'=t+\e))\, \label{regdef}
\eea
for each independent, transverse, fluctuating degree of freedom $\chi$, where  $G(x,t;x', t')$ is the appropriate to the geometry Green function. 
For fundamental strings and in flat space, the width of the flux tube increases logarithmically with the separation distance of the charges.

Here we extend these studies in the context of gauge/gravity dualities \cite{adscft1,adscft2}.
An effort in this direction has been already made in the case of strings in the fundamental representation with a
confining $AdS$ soliton background \cite{Greensite:2000cs} and with k-strings in hard wall
$AdS$ \cite{Armoni:2008sy}.\footnote{Other related constructions of connected minimal surfaces
aiming to study asymptotically the minimal area appear in \cite{Armoni:2013qda}.} In our work we approach
the problem from two different but equivalent ways. Finding the connected minimal surface between the
probe Wilson loop and the Wilson loop corresponding to the heavy Q\={Q} pair, we use certain
properties of the minimal surfaces in confining backgrounds and we prove that the logarithmic broadening
is a property of any confining gauge/gravity duality. The approximations used for our generic proof are
validated in the numerical section for the confining $D4$ Witten model, where the exact solutions of the minimal surfaces are found.

Then we look at the fluctuations of the gluonic string. Its static fluctuations can be worked out in a generic framework. 
The time dependent fluctuations are more interesting and challenging, where the generic formulas are again obtained. We begin by reviewing the time dependent fluctuations 
in flat space where we reproduce the logarithmic broadening. 
Then we move on to curved backgrounds and focus on the simplest conformal $AdS$ case, where even though the meaning of flux tube is obscure (since the gauge theory is non-confining),
an analogous quantity to the width $w^2$ can be defined.  We describe the methodology we follow and we find numerical evidence that the quantity we call 'width' in this space depends at least on one logarithmical term and is decreasing for small bound states. A toy model where the width of a string in the curved space may be obtained analytically is described in the appendix.

Our paper is organized as follows. In section 2, we compute the width of the Q\={Q} flux tube in a generic confining gauge/gravity duality, 
using the connected minimal surface between two Wilson loops, one of larger  and another of smaller radius. In section 3, we construct the exact 
numerical solution of the connected minimal surface in the D4-Witten background. In section 4, we consider fluctuations of a spatial 
string in the curved spaces, where we particularly focus on the $AdS$ space. Finally,  in section 5 we summarize and discuss our results.

%%%%%%%%%%%%%%%%%%%%%%%%%%%%%%%
\section{Tube width from circular Wilson loops}
%%%%%%%%%%%%%%%%%%%%%%%%%%%%%%%

In this section we compute the width of the tube using probe Wilson loops.
To simplify the setup we consider instead of rectangular, circular Wilson loops. This approximation is valid for the study of the
flux broadening since in the limits where the radius of the large loop is much larger than the size of the small loop and the separation distance,
the shape of the boundary loops is not expected to play the leading role in the calculation of  the width $w^2$.

%%%%%%%%%%%%%
\subsection{The setup}
%%%%%%%%%%%%%

In our analysis we use the generic metric
\be
ds^2=-g_{00}(u)dx_0^2+g_{ii}(u)dx_i^2+g_{uu}(u)du^2+dX^5~,
\ee
where we assume that it is diagonal and isotropic, so all the space metric elements are equal. The $X^5$ is the internal space
where all our configurations are kept localized. As we consider circular Wilson loops, it is advisable to make a change of coordinates
on the space-like components of the metric, $x_1=r \cos\theta~,~x_2=r \sin\theta$, giving
\be\la{metricg1}
ds^2=-g_{00}dx_0^2+g_{11} \prt{dr^2+r^2 d\theta^2 +dx_3^2}+g_{uu}du^2+dX^5~.
\ee
The boundary of the metric is taken to be at $u=0$, where  $\lim_{u\rightarrow 0} g_{11}\prt{u}=\infty$.

Large part of our analysis aims to produce the universal behavior by using certain properties of the minimal surfaces in the confining metric. At the end
we focus on a particular confining theory working numerically.

\subsection{The connected minimal surface}

We consider two circular Wilson loops at the boundary of the space with radii $R_1$ and $R_2$ $(R_1<R_2)$, with the same center, but separated
by a distance $L$ along the $x_3$ direction. The string world-sheet is parametrized by
\be\la{ansatz11}
\theta=\t~,\quad x_3=\s~,\quad r=r\prt{\s}~,\quad u=u\prt{\sigma}~,\label{WLpara}
\ee
where the static gauge is chosen along $\th$ and $x_3$.  

The Nambu-Goto action becomes
\be
S=\frac{T}{2\pi}\int d\s d\t g_{11}r\sqrt{1+r'^2+\frac{g_{uu}}{g_{11}}u'^2}:=\frac{T}{2\pi}\int d\s d\t g_{11} r\sqrt{D}
\ee
with $T$ the string tension.
Time does not appear explicitly in the action so the Hamiltonian is a constant of motion, say $-c$, specified by
\be\la{h}
c=\frac{g_{11}r}{\sqrt{D}}\, ,
\ee
that translates into the constraint equation
\be\la{h2}
1+r'^2+\frac{g_{uu}}{g_{11}}u'^2-\frac{g_{11}^2 r^2}{c^2}=0\, .
\ee
Regarding the dynamical coordinates $r$ and $u$, using eq.  \eq{h} and after some algebra we arrive at the relatively simple equations
\bea\label{r1}
&&r''-\frac{g_{11}^2 r}{c^2}=0\\
\la{u2}
&&u''\frac{g_{uu}}{g_{11}}+u'^2\frac{1}{2} \partial_u\prt{\frac{g_{uu}}{g_{11}}}-\frac{r^2 g_{11} \del_u g_{11}}{c^2}=0~.
\eea
Hence the system of equations that needs to be solved consists of eqs.
\eq{h2}, \eq{r1} and \eq{u2}. For particular class of backgrounds it can be solved analytically.

%%%%%%%%%%%%%%%%%%%%%%%%%
\subsection{Solutions for special backgrounds}
%%%%%%%%%%%%%%%%%%%%%%%%%

Let us consider the special case $g_{uu}=g_{11}$ which leads to simplifications. Equations \eq{u2} and \eq{h2} then reduce to
\bea\la{us2}
&&u''-\frac{r^2 g_{11} \del_u g_{11}}{c^2}=0~,\\
&&\la{hs2}
1+r'^2+u'^2-\frac{g_{11}^2 r^2}{c^2}=0~.
\eea
Multiplying \eq{r1} by $r$ and \eq{us2} by $u$ and adding it to \eq{hs2} we obtain
\be
\frac{1}{2}\prt{r^2}''+\frac{1}{2}\prt{u^2}''- \frac{1}{2}\frac{r^2}{c^2}\prt{4g_{11}^2+u \del_u \prt{g_{11}^2}}+1=0~.
\ee
This equation can be integrated analytically only when $g_{11}=c/u^2$, corresponding to the $AdS$ case, already done so in \cite{Olesen:2000ji}.
Other simplifications for the choice of the metric elements which lead the differential equations to analytic solutions do not seem possible. Therefore, in order to obtain analytic solutions for generic backgrounds one needs to go to certain limits in our configuration.
Thus, in the next section we consider long tubes in confining theories.

%%%%%%%%%%%%%%%
\subsection{Long flux tubes}
%%%%%%%%%%%%%%%

Concerning the limit of long flux tubes in confining backgrounds, one finds that several simplifications are possible. These come from the properties of minimal surfaces in confining spaces which are described in detail here.

The string profile along the radius $r$ of the tube lies very close to the bulk scale $u_k$ defining the tip of the confining geometry.
The projection of the surface to $(r,u)$ has almost a rectangular $\mathit{\Pi}$ shape departing from the boundary and entering into the bulk steeply.  Departing from the boundary and the small Wilson loop of radius $R_1$, the surface enters steeply in the bulk attempting to minimize its area and to shrink the radius of the tube, thus moving on the left of the $r$ axis. Then being at the bottom of the geometry the radius of the surface increases as we move to the direction of the larger Wilson loop. It extends for a distance which is almost equal to the separation of the radii on the boundary $R_2-R_1$, and then just before the radius reaches the $R_2$ value returns steeply to the boundary.

The projection of the world-sheet along the $(u,x_3)$ of the two Wilson loops is expected to lead again to an inverted $\Pi$
shape with a mostly rounded corner on the side of the small loop.  The minimal surface at the side of the large radius Wilson loop goes deeper in the bulk to reach the smaller area elements, than the side that is closer to the small loop. The turning point of the string along the direction $x_3$
is not in the middle of the string but lies closer to the side of the large Wilson loop. A large portion of the string world-sheet lies around
the turning point and it is this region that gives the dominant contribution to the minimal surface. 
The qualitative picture of the minimal surface solution described here is the same for any background and will be later verified numerically, by finding the particular solution of the surface in the D4 -Witten model.

To proceed with the calculation let us define the two quantities
\be
h(u):= \frac{g_{11}^2}{c^2},\hskip 1cm f(u) := \frac{g_{uu}}{g_{11}}~,
\ee
that allow us to write the system of equations in the more compact form
\begin{eqnarray}\la{systema1}
&& r'' - h r =0~,\nonumber\\
&& 2u'' +(u')^2 \partial_u \left(\ln f\right) - r^2 \frac{\partial_u h}{f} = 0~,\\\nonumber
&& (r')^2 + f (u')^2 = h r^2 -1\, .
\end{eqnarray}
At the bottom of the world-sheet where $u=u_0$ and $u'=0$ we can determine its shape by solving the system
\begin{eqnarray}
 r'' - h r = 0 ~,\qquad (r')^2 +1 = h r^2~,
\end{eqnarray}
yielding to
\be
r=\frac{e^{s\sqrt{h}\prt{c_1+\s}}\prt{e^{-2 s \sqrt{h} \prt{\s +c_1}}+ h}}{2 h}\la{tip1}\, ,
\ee
where $c_1$ can be taken to zero without loss of generality and we choose the sign $s=-1$.

The radius of the flux tube at its endpoints is specified by the boundary conditions
\be\la{bbcc1}
r(0) = R_1,\qquad r(L) = R_2~.
\ee
Since the string world-sheet dives directly into the bulk and extends down to the tip of the geometry from where it takes its main contributions, it is natural to assume that the radius of the
flux tube in the bulk around $\s=0$ and $\s=L$ approaches the radius of
the string on the boundary. And since the string lies on the tip of the horizon we can apply these boundary conditions to the function $r\prt{\s}$ obtained in \eq{tip1}, to get the leading contributions
\be\la{radii1}
R_1=\frac{1}{2}\prt{1+\frac{1}{h}}~,\qquad R_2=\frac{1}{2}\prt{e^{- \sqrt{h} L}+\frac{1}{h}e^{ \sqrt{h} L}}~,
\ee
where $h$ now is computed for $u \simeq u_k$, close to the tip of the confining geometry. 
Notice that for large $L$ a logarithmic dependence of the length on the ratio of the radii appears as
\be\la{largel1}
L=\frac{1}{\sqrt{h}} \log \frac{R_2}{R_1}~.
\ee
Minimizing the action and using the Virasoro constraint we obtain
\be
S=-\sqrt{\l}\frac{g_{11}\prt{u_0}^2}{c}\int_0^L d\s r^2\, .
\ee
Integrating the action and using \eq{radii1} and \eq{largel1} we obtain
\be
S= \frac{g_{11}\prt{u_k}}{2}\prt{\frac{L}{\sqrt{h}}+R_2^2-R_1^2+\frac{1}{2}\prt{1-\e^{-2 \sqrt{h} L}}}\simeq
\s\prt{\frac{L^2}{\log\frac{R_2}{R_1}}+R_2^2-R_1^2}\, ,
\ee
where we have used that $u_0\simeq u_k$ and taken $R_2\gg R_1$.
Notice that for the string tension we have substituted $g_{11}\prt{u_k}/2$.
The energy of the string obtained for confining backgrounds, in this limit, approaches the one obtained when the string is placed in flat space.

Inserting the area of the minimal connected surface in \eq{wdef1} for the transverse directions, we obtain
\be
w^2\simeq\log\frac{R_2}{R_1}~.
\ee
Evidently the width of the tube has a logarithmic dependence on the radius $R_2$, i.e. on the inter-quark distance.
Notice that the position of the small probe loop does not affect the result, as expected.

%%%%%%%%%%%%%%%%%%%%%%%%%%%%%%%%%%%%
\section{Flux tube and connected minimal surfaces: A numerical treatment}
%%%%%%%%%%%%%%%%%%%%%%%%%%%%%%%%%%%%

In this section we present a numerical treatment of the connected minimal surface between the Wilson loop
associated with the heavy quark bound state and the probe Wilson loop. We remark that the tube-like surfaces that we plot
do not correspond to the actual QCD flux-tubes, but result from the definition \eq{wdef1}. We obtain the exact minimal surface numerically,
and plot the projections of the radial
profile of the string $u\prt{\s}$ and the cross section $r\prt{\s}$ in terms of the distance $\s=x_3$.
The findings of the numerical analysis show the validity of the analytical approximations made in the previous section.

The theory we choose is the confining D4 Witten model  \cite{Witten:1998zw}, with metric written in the form \eq{metricg1}
\bea
&&ds^2=\prt{\frac{u}{R}}^{3/2}
\prt{-dt^2+ dr^2+r^2 d\theta^2 +dx_3^2+f(u)dx_4^{2}}
+\prt{\frac{R}{u}}^{3/2}
\left(\frac{du^2}{f(u)}+u^2 d\Omega_4^2\right),\label{d4m}
\nonumber\\
&&e^\phi= g_s \left(\frac{u}{R}\right)^{3/4},\quad
~f(u)=1-\frac{u_k^3}{u^3} \ ,\quad R^3=\pi g_s N_c l_s^3 ~,
\label{D4m1}
\eea
where $d\Omega_4^2$ is the metric of a unit $S^4$ of radius $R$ on which the string is localized.
The spatial compactification is along $x_4$ and the tip of the geometry is denoted by $u_k$. The string tension in this model is
$1/\prt{2\pi\a'} \left(u_k/R\right)^{3/2}$.

The Wilson loops are placed in this background with the parametrization \eq{WLpara} with
the boundary conditions of the previous section. Here we solve the system of equations \eq{systema1} numerically, with no approximations.
The input to the system is a choice for the quantities $r_0$, $u_0$, $u'\prt{\s_0}=0$ and the value of $r'\prt{\s_0}$.
The derivative of $r$ at the turning point $\s_0$ is obtained by solving \eq{h2} and using the values chosen already for the other parameters.
The solution to the differential equations gives two functions of the form $r\prt{s}$ and $u\prt{s}$.

The setup for the minimal surface is the following: The boundary of the space is at infinity as can be seen from \eq{D4m1}.
We place the small Wilson loop with radius $R_1$ at smaller distance from the origin along the direction of the tube $x_3$, and the
large Wilson loop with radius $R_2$ at larger distance. The order of scales is $R_2>R_1$.
The particular numerical values of the scales will be chosen to ease the presentation. The area of the minimal surfaces is regularized by either
inserting a cut off close to the boundary or by substracting the infinite
parts of the action \cite{Drukker:1999zq,Chu:2008xg}.

\begin{figure}[!ht]
\begin{minipage}[ht]{0.5\textwidth}
\begin{flushleft}
\centerline{\includegraphics[width=80mm]{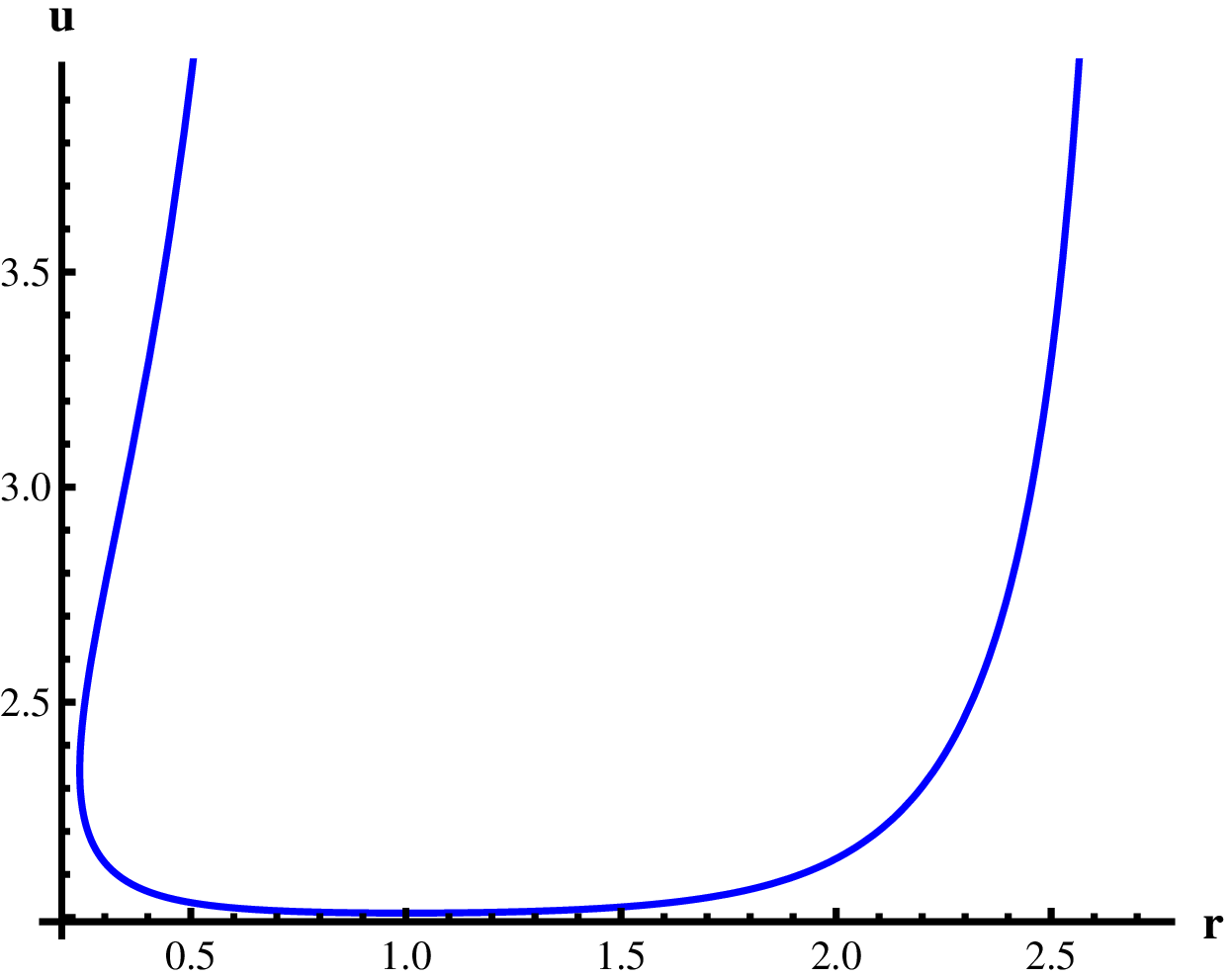}}
\caption{\small{The profile of the minimal surface along the holographic direction $u$ and its radius $r$.
The shape of the curve resembles an inverted $\mathit{\Pi}$, while the bottom of the string is moved to the
left where the smaller loop is located. The minimum radius of the tube is inside the bulk and closer to the small loop.
\vspace{0.9cm}}}\label{ur}
\end{flushleft}
\end{minipage}
\hspace{0.3cm}
\begin{minipage}[ht]{0.5\textwidth}
\begin{flushleft}
\centerline{\includegraphics[width=80mm]{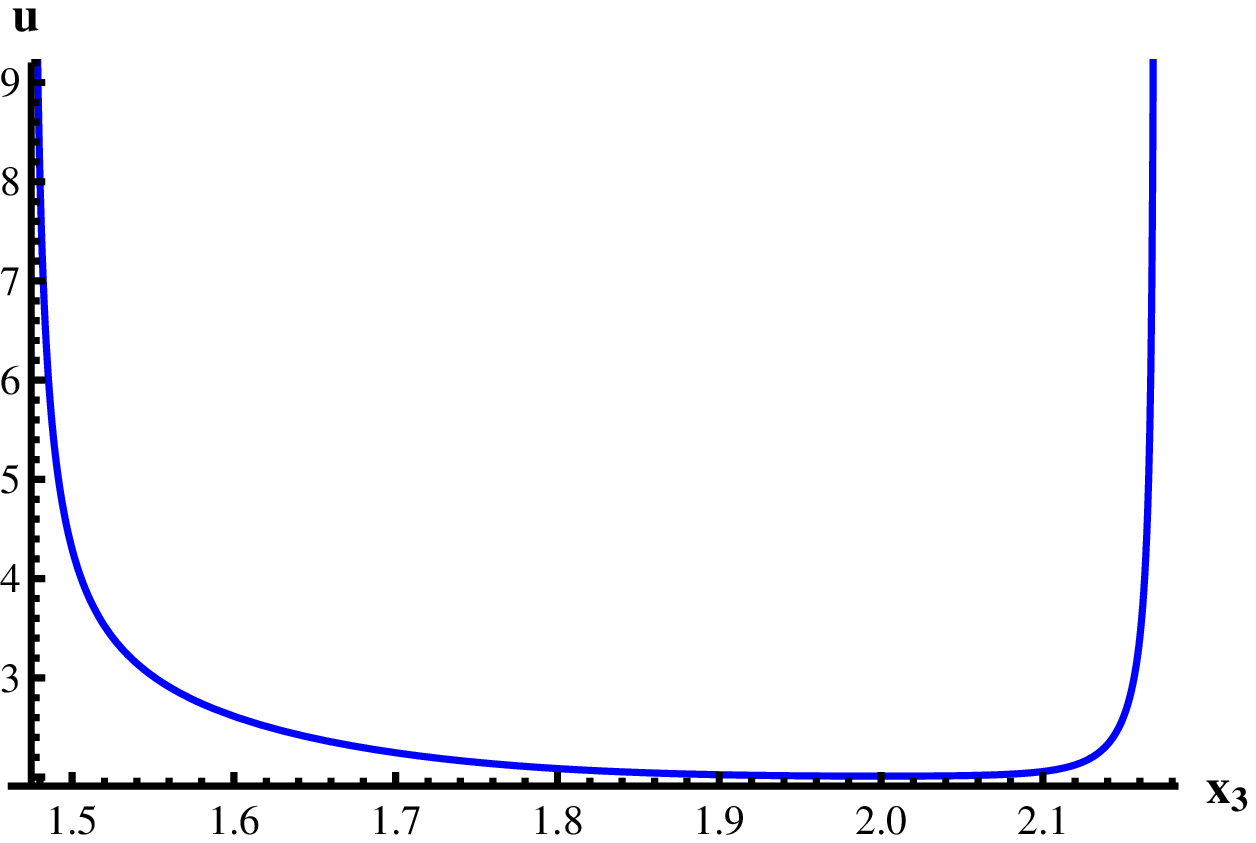}}
\caption{\small{The minimal surface projected along $u\prt{x_3}$. For large values of $x_3$ where the larger
Wilson loop is placed, the surface goes slightly deeper in the bulk, in order to minimize its area. The shape of the curve approaches
again that of an inverted $\Pi$, with the side closer to the small Wilson loop slightly rounded.
\vspace{0cm}}}\label{ux3}
\end{flushleft}
\end{minipage}\vspace{-1.1cm}~
%\end{figure}
%\begin{figure*}[!ht]
\newline
\centerline{\includegraphics[width=80mm]{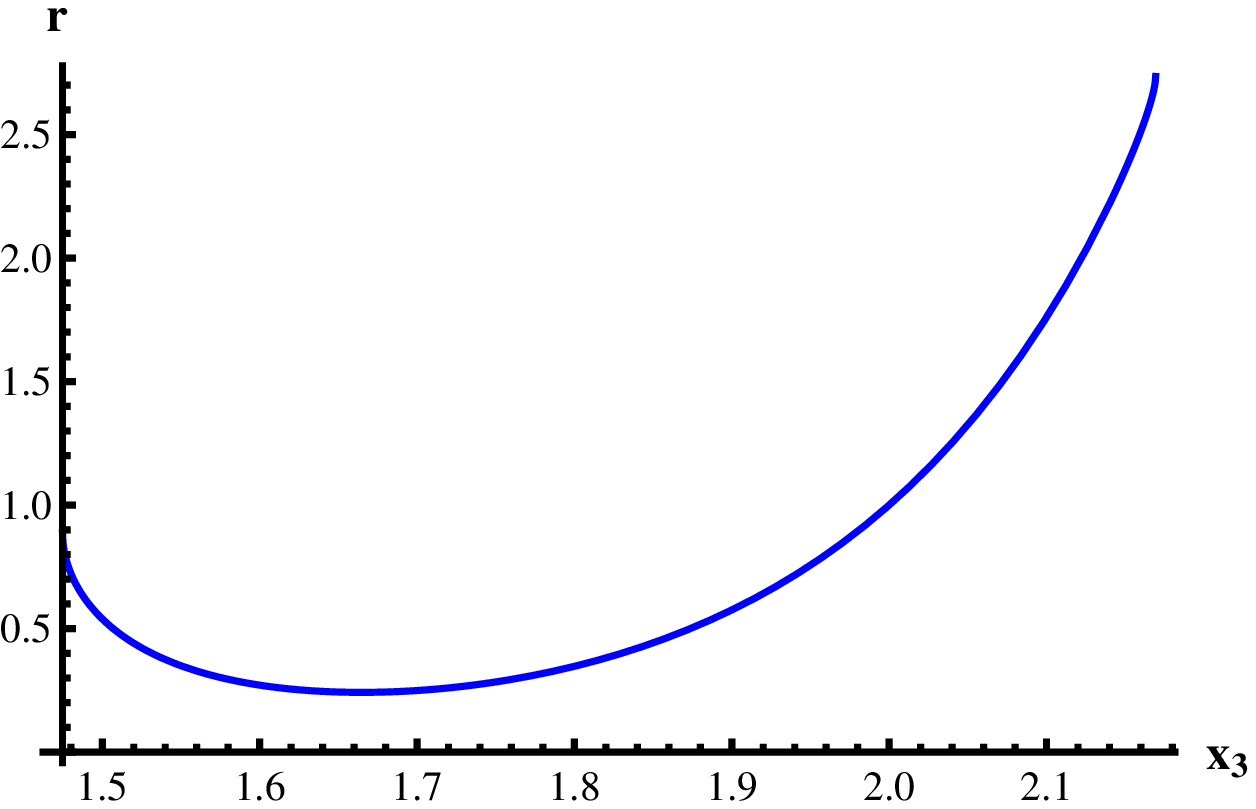}}
\caption{\small{The radius of the string along the spatial extension of the tube. It is not a monotonic function and appears to have minimum close to the small Wilson loop.
\vspace{0cm}}}\label{rx3}
\end{figure}

In Figure \ref{ur} the solution for $u\prt{r}$ is plotted. We observe the almost $\mathit{\Pi}$-like shape of the string described in the previous section,
where most of the world-sheet lies along the tip of the geometry. As the string enters into the bulk, the radius of the tube decreases at both ends.
As expected, the smaller tube radius is observed on the left where the smaller Wilson loop is located.
Notice that the minimum radius is inside the bulk and not on the boundary.

In Figure \ref{ux3} the function $u\prt{x_3}$ is shown. The rounded from one side, $\Pi$ shaped Wilson loop is presented. The minimal surface
prefers to go deeper into the bulk to smaller area elements at the side of large radii in order to minimize its area more effectively.
That is why the minimum value along the holographic direction is closer to the large Wilson loop. Around this point the string lies
almost flat for a long range. Moreover, this is the region from where the major contributions to the action come.

In Figure \ref{rx3} the solution for $r\prt{x_3}$ is presented, representing the radius of the tube along the $x_3$ direction.
Starting from the small loop the radius decreases along the tube. The minimal value of the radius is
closer to the small loop, after the turning point it increases until it becomes equal to the one of the large tube.
%%%
\begin{figure*}[!ht]
\begin{minipage}[ht]{0.5\textwidth}
\begin{flushleft}
\centerline{\includegraphics[width=80mm]{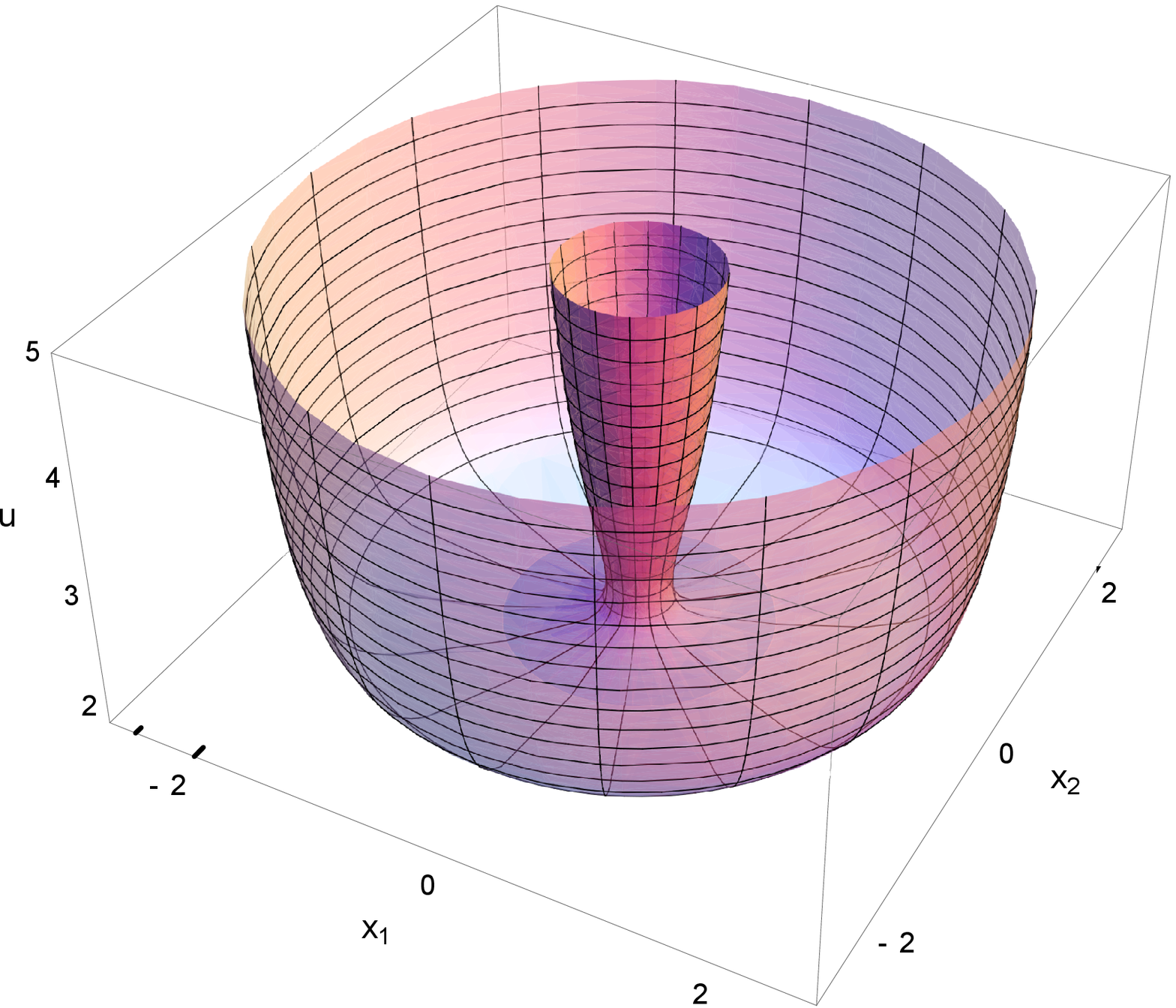}}
\caption{\small{The two dimensional surface of the radius of the tube as it enters into the bulk. Notice that the minimum of the
radius of the tube occurs very close to the minimum holographic radial distance. The increase of the radius from its minimum
value to the value of the radius $R_2$ of the larger Wilson loop happens almost totally around the tip of the geometry $u_k$.
\vspace{0cm}}}\label{pru}
\end{flushleft}
\end{minipage}
\hspace{0.3cm}
\begin{minipage}[ht]{0.5\textwidth}
\begin{flushleft}
\centerline{\includegraphics[width=80mm]{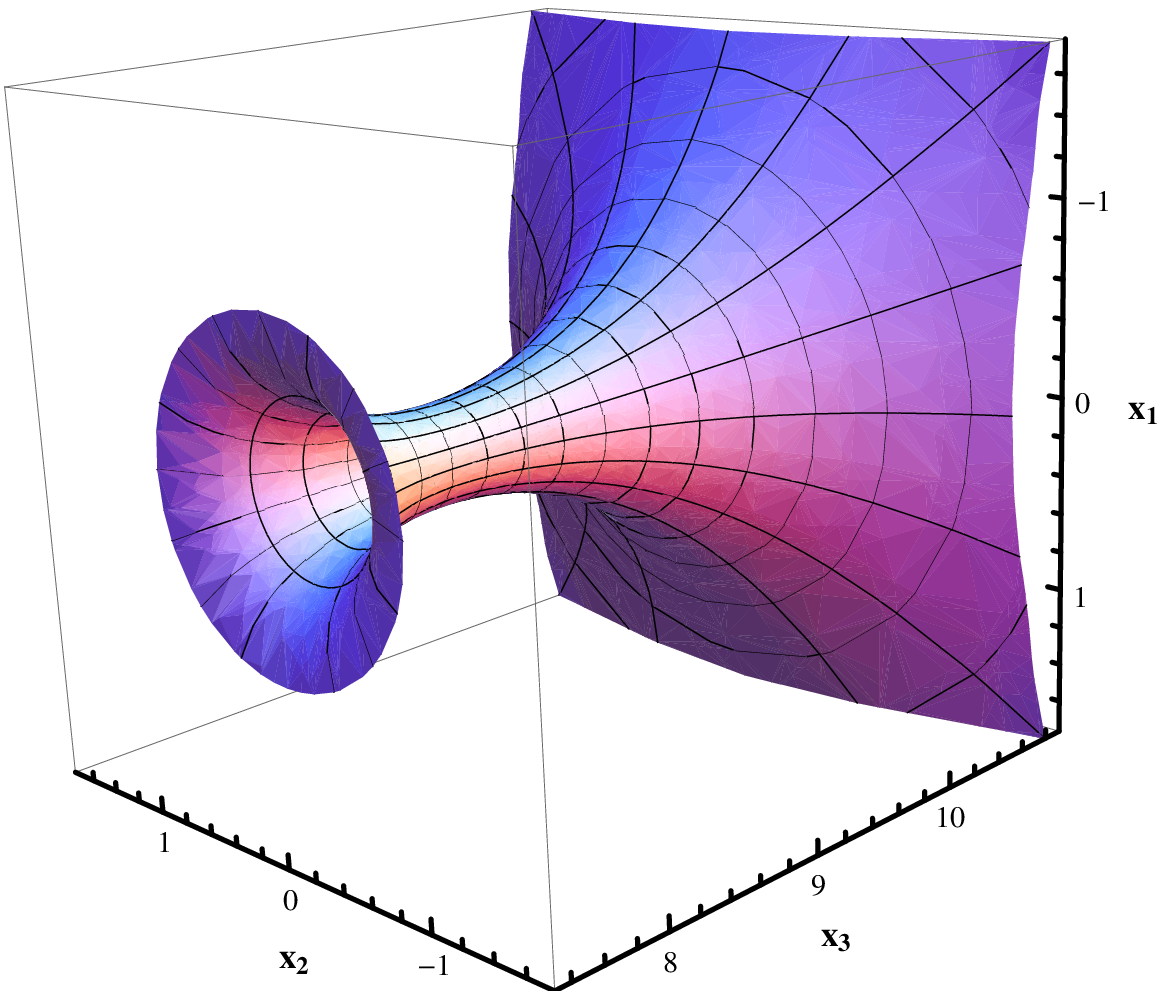}}
\caption{\small{The minimal surface in the spatial dimensions. The small Wilson loop is placed at small $x_3$ and the large one on the other side.
When the surface departs from the Wilson loops its radius is decreasing considerably. The spatial $x_3$ extension of the tube in the particular plot is five times enhanced for presentation purposes.
\vspace{0.1cm}}}\label{prx}
\end{flushleft}
\end{minipage}
\end{figure*}
%%%

In Figure \ref{pru} the two dimensional surface $(r\prt{u},t)$ of the tube entering the bulk is presented in $x_i$ coordinates, used to bring the metric in the form \eq{metricg1}.
The $x_3$-dimension of the tube is not depicted. The inner funnel corresponds to the Wilson loop of radius $R_1$ and as the
surface enters into the bulk its radius becomes smaller. The inner funnel radius has reached already its smallest value before
it reaches the tip of the geometry at the minimal holographic distance. The radius increases rapidly from its minimal value towards $R_2$
close to $u_0$. Having approached a value close to $R_2$, the surface returns into the bulk.
The reason for this is that the main contribution to the area of the minimal surface comes from the region around $u_0$.

In Figure \ref{prx} the two dimensional surface $(r\prt{x_3},t)$ of the tube is shown in the $x_i$ coordinates. Along the spatial direction $x_3$ the radius of the tube
decreases until it reaches a minimal point and increases again when the loops are approached. The representative values we
have chosen for the particular plots are $c=0.5,~u_0/u_k=1.005,~r_0=1$ that then fix
$R_1=0.89,~R_2=2.74$ for the radii and $L_1\prt{1}=1.48,~ L_2\prt{2}=2.17$ for the position of the loops along the $x_3$ direction.

%%%%%%%%%%%%%%%%%%%%%%%%%%%%%%%%%%
\section{Fluctuations of the spatial string}
%%%%%%%%%%%%%%%%%%%%%%%%%%%%%%%%%%

In flat space the effective string description shows that the broadening of the flux tube at finite temperature,
below the deconfinement phase transition, becomes linear.
The linear broadening is supported also by Monte Carlo simulations in certain simple models \cite{Allais:2008bk}.
This means that there may be another order parameter characterizing confining gauge theories. From the point of view of string theory on the other hand it may be also used to distinguish
confining backgrounds, as the expectation value of the Wilson loop does. The quantity $w^2$ in eq. \eq{regdef} is well defined not only for confining theories but for a larger class of backgrounds.

The simplest object for which we can compute its fluctuations is a spatial string.
Our bulk order parameter will therefore be $\d x_2(\s_0)$,
with $x_2$ one of the boundary, transverse degrees of freedom and $\s_0$
the world-sheet parameter specifying the middle of the string.

%%%%%%%%%%%%%%%%%
\subsection{Static fluctuations}\label{stat:1}
%%%%%%%%%%%%%%%%%

Let us consider the spatial string
\be
t=\t~,\quad x_1=x_1\prt{\s}~, \quad x_2=0~, \quad u=\s~,\label{spatial}
\ee
that leads to the action
\be\la{action12}
S=\frac{T}{2\pi}\int du \sqrt{g_{00}\prt{g_{uu}+g_{11} x_1'\prt{\s}^2} }~
\ee
and the equation of motion
\be\la{length1a}
x_1'\prt{\s}=\frac{\sqrt{g_{00_0}g_{11_0} F}}{g_{00}g_{11}}~,
\ee
with $g_{11_0}:=g_{11}\prt{\s_0}$, $g_{00_0}:=g_{00}\prt{\s_0}$ and
\be
F:=\frac{g_{00}^2g_{uu}g_{11}}{g_{00}g_{11}-g_{00_0}g_{11_0}}\, .
\ee
Perturbing
\be
x_1=x_1\prt{\s}+\d x_1 \prt{\s}~,\quad x_2=\d x_2\prt{\s}~,
\ee
one sees that the first order contributions cancel out \cite{Avramis:2007mv} while to second order the contribution of the fluctuations is
\be
S_2=\int d\s \prt{\frac{g_{00}^2 g_{uu} g_{11}}{2 F^{3/2}}\d x_1'^2+\frac{g_{00}g_{22}}{2 \sqrt{F}}\d x_2'^2}\,
\ee
and the resulting linearized equations of motion are
\bea\la{eq1a}
\frac{d}{d\s}\prt{\frac{g_{00}^2 g_{uu} g_{11}}{F^{3/2}}\d x_1'}=0\, , \qquad \frac{d}{d\s}\prt{\frac{g_{00} g_{22}}{F^{1/2}}\d x_2'}=0~.
\eea
Since we care for the transverse fluctuations $\d x_2$ and evidently the two types of fluctuations
decouple from each other, we can set consistently the longitudinal part to zero.

The two equations then that govern the string configuration and the fluctuations are
\bea\la{eqx1}
x_1\prt{\s}-x_1\prt{\s_0} &=& \sqrt{g_{00_0}g_{11_ 0}}\, I_1, \hskip 1cm I_1(\s; \s_0) = \int_{\s_0}^\s d\s \frac{\sqrt{F}}{g_{00} g_{11}}~,\nonumber\\
\d x_2\prt{\s} -\d x_2\prt{\s_0} &=& -c_1 I_2, \hskip 1.85cm I_2(\s; \s_0) = \int_{\s_0}^\s d\s \frac{\sqrt{F}}{g_{00} g_{22}}~.
\eea
For spatially isotropic backgrounds $g_{11}=g_{22}$ and thus $I_1=I_2\equiv I$ and $x_1\prt{\s_0}$ is fixed by symmetry to be the middle point of the string.
The solution is determined by the boundary conditions
\bea\label{bbbc}
 \frac{L}{2}=\sqrt{g_{11_ 0}g_{00_0}}\, I(\infty ; \s_0) ~, \qquad
\d x_2\prt{\s_0}= c_1  I(\infty ; \s_0)  ~,
\eea
that yield for our quantity of interest the expression
\be
\d x_2(\s_0) = \frac{c_1}{2 \sqrt{g_{00_0}g_{11_ 0}}}\, L\, ,\label{s1}
\ee
where $g_{11_ 0}$ should be determined by substituting $\s_0(L)$ from the first expression of \eq{bbbc}. 
The validity of the solution depends of course on the continuity conditions of the string in the turning point and should be checked for each background.

\textbf{Particular backgrounds}

The \eq{eqx1} can solved for two particular backgrounds, a conformal and a confining.

The metric elements for the $AdS$ background are given by $g_{11}=g_{22}=u^2/R^2$ and $g_{uu}=R^2/u^2$, where $R$ is the radius of the $AdS$ space where we set it equal to the unit and we recover it only in the final result.
Then
\be
I(\s;\s_0) = \int_{\s_0}^{\infty} d\s \frac{1}{\s^2\sqrt{ (\s^4-\s_0^4)}} = \frac{\sqrt{\pi } \Gamma \left(\frac{3}{4}\right)}{\s_0^3 \Gamma \left(\frac{1}{4}\right)}~.
\ee
The length $L$ on the boundary is then given by \cite{Maldacena:1998im}
\be
L=\frac{2 \sqrt{\pi }  \Gamma \left(\frac{3}{4}\right)}{\s_0 \Gamma \left(\frac{1}{4}\right)}\, .
\ee
The metric elements for the confining $D4$-brane background are $-g_{00}=g_{11}=g_{22}=u^{3/2}$ and $g_{uu}=1/\prt{u^{3/2} f}$, where $f=1-u_k^3/u^3$,
where we have set the radius $R$ of the sphere equal to one and recover it for the final result. The length of the string is given by
\be
L = 2\s_0^{3/2} \int_{\s_0}^\infty \frac{d\s}{\sqrt{(\s^3-\s_0^3) (\s^3-u_k^3)}}
\simeq\frac{ -2 \log\left(\s_0-u_k\right)+ const.}{3 \sqrt{u_k}}~,
\ee
where the constants containing terms from the boundary, are omitted. The turning point of the string is now determined by $c^2=\sqrt{g_{00_0} g_{11_0}}$
and the analysis is similar to \cite{Giataganas:2011nz,Giataganas:2011uy}, although in a different gauge. Then we obtain that
\be\label{uo11}
\s_0\simeq u_k\left(1+ e^{-3 L \sqrt{u_k}/2} \right)~ .
\ee

%%%%%%%%%%%%%%%%%%%%%%
\subsection{Time dependent fluctuations}
%%%%%%%%%%%%%%%%%%%%%%

In this section we consider time-dependent fluctuating strings in curved backgrounds. The expectation value of the square transverse deviations of the string with periodic boundary conditions in Euclidean time, has been studied extensively in \cite{Gliozzi:2010zv,Gliozzi:2010zt}  and is defined as explained in \eq{regdef} 
\bea
w^2(x) = \langle \chi(x,t)^2 \rangle = \lim_{\e\to 0}G(x,t;x'=x+\e,t'=t+\e))~,\label{regdef2}
\eea
for each independent, transverse, fluctuating degree of freedom $\chi$, with $G(x,t;x', t')$ the appropriate to the geometry Green function.
The coordinate $x$ is some space-like world-sheet coordinate while $t$ is the corresponding time-like coordinate.

With time dependence built in as in \cite{Avramis:2007mv}, the action of the perturbations is computed from \eq{action12} to give
\be
S_2=\frac{1}{2}\int d\s \prt{\frac{g_{00}^2 g_{uu} g_{11}}{ F^{3/2}}\d x_1'^2+\frac{g_{00} g_{22}}{\sqrt{F}}\d x_2'^2-\frac{g_{11} g_{uu}}{\sqrt{F}}\d \dot x_1^2- \frac{g_{22}  \sqrt{F}}{g_{00}}\d \dot  x_2^2}\, ,
\ee
leading to the equations of motion
\bea
&&\frac{d}{d\s}\prt{\frac{g_{00}^2 g_{uu} g_{11}}{ F^{3/2}}\d x_1'} -\frac{g_{11} g_{uu}}{\sqrt{F}} \d\ddot x_1=0\, , \\
&&\frac{d}{d\s}\prt{\frac{g_{00} g_{22}}{\sqrt{F}}\d x_2'}-\frac{g_{22} \sqrt{F}}{g_{00}} \d\ddot x_2=0~.
\eea
We will restrict ourselves to the isotropic case $g_{11}=g_{22}$.
The solutions for the string worldsheet are the ones we have already obtained while the solution
for the perturbed string need to be reexamined due to the additional fluctuations.

\subsubsection{The flat background}
%%%%%%%%%%%%%%%%%%

According to \eq{regdef} we need to calculate the Green function of a Laplace equation on a two dimensional plane with one Dirichlet and one periodic boundary condition. The Green function of this cylinder with dimensions $L\times \b$ ($\b$ can be thought of as a Euclidean time) can be mapped to that of a torus with size $\tilde{L}\times \b$ where $\tilde{L}=2 L$. The Green function of the Laplace equation with toroidal boundary conditions is given by \cite{Gliozzi:2010zt}  
\be\la{greent}
G\prt{x,t}=\frac{t\prt{t-\b}}{2 \s\b \tilde{L}}+\frac{1}{2 \pi \s}\sum_{n=1}^\infty \cos\prt{\frac{2 \pi n x}{\tilde{L}}}\frac{1}{n\prt{1-q^n}}\prt{\exp{\frac{-2 n \pi t}{\tilde{L}}}+q^n \exp{\frac{2 n \pi t}{\tilde{L}}}}+K~,
\ee
where  $q=e^{2i\pi \tau}$, $\tau=i\b / \tilde{L}$, $K=\b/(12\s \tilde{L})+\log\eta\prt{\t}/ \prt{\pi\s}$ and $\s$ is a string tension, not to be confused with the world-sheet order parameter.

The cylinder Green function is then obtained from that of the torus by
\bea\nn
&&G_c\prt{x,t;x',t'}=G_T\prt{z-z'}-G_T\prt{z+\bar{z}'}\\
&&\hspace{2.4cm}=\frac{1}{\pi\s}\sum_{n=1}^\infty \sin\frac{\pi n x}{L} \sin\frac{\pi n x'}{L}\frac{e^{-\pi n \prt{t-t'}/L}+e^{-\pi n \prt{\b-t+t'}/L}}{n\prt{1-q^n}}~.\la{flatb1}
\eea
The width is ultraviolet divergent and is regularized with the point-splitting method,
where the primed coordinates are assumed to be $\e$ away from the non-primed coordinates,
in the limit $\e\to 0$ giving \cite{Gliozzi:2010zv,Gliozzi:2010zt,Gliozzi:2010jh}
\be\la{widthw}
w^2\prt{L/2}=\frac{1}{2\pi\s}\log{\frac{L}{L_0}}+\frac{1}{\pi\s}\log \eta\prt{ 2\t}\, ,
\ee
with $L_0$ proportional to $\e$. In the computation one finds essential the property
\bea
\sum_{n=1}^\infty\frac{n^{-1}q^n}{1-q^n}= -\sum_{n=1}^\infty\log\prt{1-q^n}=-\log\phi\prt{\t}~,
\eea
where the function $\phi\prt{\t}$ is the Euler function, related to the Dedekind $\eta$-function by $\phi\prt{\t}=\eta\prt{\t} q^{-1/24}$.

Equation \eq{widthw} has two terms. When $\b\gg L$ the leading term is the logarithm, and the logarithmic broadening is obtained. When  $L\gg \b$ the width becomes instead
\be
w^2(L/2) = \frac{1}{2\pi \s} \log \frac{\b}{4L_0} +\frac{1}{4\b \s} L~,
\ee
and the leading term becomes linear.
The fluctuations in the former limit are more interesting since are closer to the Wilson loop setup in terms of the limits used in holography.

%%%%%%%%%%%%%%%%%%%%
\subsubsection{Curved backgrounds}
%%%%%%%%%%%%%%%%%%%%

We still assume periodic time and a Euclidean signature on the worldsheet and introduce the notation
\be
\d x_2(\s,t) \equiv \chi (\s,t)\, .
\ee
In order to use the AdS metric, we are working in the low temperature limit $L T\ll 1$. The equation per independent fluctuation reduces to
\bea
\partial_\s \left(\frac{g_{00} g_{11}}{\sqrt{F}} \partial_\s\chi(\s,t)\right) - \frac{g_{11} \sqrt{F}}{g_{00}} \, \partial_t^2 \chi(\s,t)=0~,\label{Dmain}
\eea
where $g_{11}$ and $F$ are functions of $\s$. We express everything in parameters $x = {\s}/{\s_0}$ with $x\in [1,\infty)$
and $\tilde{t}=t \s_0^{-s/2}$ and rewrite the above differential equation as
\be
(p \chi')' - r {\ddot \chi} = 0\label{timeSL}~,
\ee
where
\bea
p = \frac{g_{00} g_{11}}{\sqrt{F}}~,\qquad r =\frac{g_{11} \sqrt{F}}{g_{00}}~,
\eea
are functions of $x$ only, the prime denotes
differentiation with respect to $x$, the dot differentiation with respect to $\tilde{t}$ and $s$ is the total scaling dimension of the equation.
The periodicity of the time direction $\b$ then becomes ${\tilde \b} = \b \s_0^{-s/2}$.

We are therefore looking for the Green function $G$ that satisfies the equation
\be
(p G')' - r {\ddot G} = \d(\tilde{t}-\tilde{t}')\, \d(x-x')\, \label{GreenEq1}
\ee
and obeying the jump condition
\be
- r \left( {\dot G}(\tilde{\b}) - {\dot G}(0)\right) = \d(x-x')\, ,
\ee
obtained by integrating eq. \eq{GreenEq1} over time.
Next we define the expansions
\bea
\chi(x,\tilde{t}) &=& \sum_n c_n(\tilde{t}) w_n(x) \label{com1}~,\\
\rho(x,\tilde{t}) &=& \sum_n \rho_n(\tilde{t}) w_n(x),\hskip 1cm \rho_n(\tilde{t}) = \int_1^\infty \rho(x',\tilde{t}) w_n(x') r(x') dx'~,\label{com2}
\eea
where
\be
w_n(x) = N_n u_n(x),\hskip 1cm N_n = \frac{1}{\sqrt{\int_1^\infty u_n(x)^2 r(x) dx }}~,\label{normf1}
\ee
is a set of complete, orthonormal functions on $[0,\infty)$ satisfying
\bea
\int_1^\infty dx\, w_n(x) w_m(x) \, r(x) = \d_{n m}~,\qquad
\sum_n w_n(x)\, w_n(x') r(x') =  \d(x-x')~.
\eea
We have introduced the auxiliary source function $r(x) \rho(x,\tilde{t})$ in order to compute the Green function. Inserting on the
right hand side of \eq{timeSL} the source we obtain (summation over $n$ is implied)
\bea
c_n (p w_n')' - r {\ddot c}_n w_n = r(x) \rho_n(\tilde{t}) w_n(x)~.\label{eqxt}
\eea
Requiring $w_n$ to solve
\be
(p w_n')' - \Omega_n^2\, r w_n = 0~,\label{unnor}
\ee
eq. (\ref{eqxt}) is satisfied if
\bea
{\ddot c}_n - \Omega_n^2 c_n = - \rho_n\, . \label{SLt3}
\eea
The solution to the above, with the periodic boundary conditions in the time direction, is
\bea
c_n(\tilde{t}) = \int_0^{\tilde\b} G_{{\prt{\tilde{t}}}_n} (\tilde{t}-\tilde{t}') \left(-\rho_n(\tilde{t}')\right) d\tilde{t}'= \int_0^{\tilde\b} \frac{ e^{-\Omega_n (\tilde{t}-\tilde{t}')} + e^{-\Omega_n ({\tilde\b} - (\tilde{t}-\tilde{t}'))} }{2 \Omega_n\prt{1-e^{-\Omega_n {\tilde\b}}}} \rho_n(\tilde{t}') d\tilde{t}'\, ,
\eea
where the Green function $G_{{\prt{\tilde{t}}}_n} (\tilde{t}-\tilde{t}')$ depends only on the dimensionless time.
Then, using eq. \eq{com1} we have
\be
\chi(x,\tilde{t}) = \int_0^{\tilde\b} \int_1^\infty G(x,\t; x',\tilde{t}')\, \rho(x',\tilde{t}') r(x')\, dx' d\tilde{t}'~,
\ee
with full Green function
\be
G(x,\tilde{t}; x',\tilde{t}') = \sum_n  \frac{ e^{-\Omega_n (\tilde{t}-\tilde{t}')} + e^{-\Omega_n ({\tilde\b} - (\tilde{t}-\tilde{t}'))} }{2\Omega_n\prt{1-e^{-\Omega_n {\tilde\b}}}}\, w_n(x)\, w_n(x') \, .\label{finalGreen}
\ee
In terms of dimensionful time this is
\be
G(x,t; x',t') = \sum_n  \frac{ e^{-\Omega_n (t-t')\s_0^{-s/2}} + e^{-\Omega_n ({\b} - (t-t'))\s_0^{-s/2}} }{2\Omega_n\prt{1-e^{-\Omega_n {\b}\s_0^{-s/2}}}}\, w_n(x)\, w_n(x')~.
\ee
Notice that for the $AdS$ metric analysis presented in the following section, $s=-2, \s_0\sim 1/L$ and for the toy example presented in 
Appendix \ref{app:QQtoy}, $s=+2, \s_0\sim L$, the dimensionless time parameter is $\t=t/L$ and the Green function for both cases becomes
\be
G(x,t; x',t') = \sum_n  \frac{ e^{-\Omega_n (t-t')/L} + e^{-\Omega_n ({\b} - (t-t'))/L} }{2\Omega_n\prt{1-e^{-\Omega_n {\b}/L}}}\, w_n(x)\, w_n(x') \, ,
\ee
with $w_n$ satisfying eq. \eq{unnor}.

%%%%%%%%%%%%%%%%%%%%%%%%%%%%%%%%%%%%%%
\subsection{The width of the fluctuation}\label{s:2point}
%%%%%%%%%%%%%%%%%%%%%%%%%%%%%%%%%%%%%%

Working with the metric elements $g_{11}=\s^2$, $g_{uu}=1/\s^2$ we have
\bea
s = -2~,\quad r = \frac{x^2}{\sqrt{x^4-1}}~,\quad p &=& x^2 \sqrt{x^4-1}~.
\eea
Writing \eq{unnor} in the form of
\be
y'' +\frac{p'}{p} y' - {\Omega_n^2}\frac{r}{p} y = 0 \label{Legendre1}
\ee
the equation for the AdS Wilson Loop becomes
\be
y'' +\frac{2\prt{1-2x^4}}{x\prt{1-x^4}} y' + {\Omega_n^2 }{\prt{x^4-1}} y = 0\, . \label{eq22}
\ee
We need to find the Green function of this equation and perform the point splitting around $x=1$.
An analytic computation does not seem to be possible, but a
numerical version with certain approximations may be done.

To begin, we can solve numerically eq. \eq{eq22} in order to specify the solutions $w_n$ up to normalization constants.
Then using the formalism of the previous section yielding the Green function \eq{finalGreen},
we can specify numerically the Green function of \eq{eq22}.  The point-splitting is also done numerically
and taken close to the value $x=1$, where the result is integrated over all positive values of $\o$.
Before listing the results we mention a few shortcomings of the numerics. Although we
can go infinitely close to $x=1$ we can not take the exact value of unity and therefore we are forced to introduce a cut-off $x_{\rm min}=1+\e_2$.
The integration in all positive values of $\o$ should also start with a non-zero value $\o_{\rm min}=\e_1\ll 1$ and arbitrarily close to zero.
Therefore, in the numerics we have introduced these two cut-offs even though the normalized result seems to depend weekly
on their exact values as long as they are small enough.
It should be finally mentioned that we work with large numbers, as expected in these limits, that could be the source of numerical uncertainties.

Our numerics indicate that the leading contributions come from the time dependent part of the Green function.
Moreover, we find that the width is linear in $L/\prt{\beta \e_1}$ to a good approximation. This is a term that
exists also in the width of the toy model configuration presented in Appendix A and appears in equation \eq{wlog2}.
The cut-off independent term that we find to a good accuracy is the logarithm $\ln (L/\b)$, also found in the configuration
of Appendix in \eq{w222}.
Therefore, the total fitted expression reads
\be
w^2\simeq c_1 \frac{L}{\b \e_1}-c_2 \ln \frac{L}{\b}+ F\prt{\e_1,\e_2,\frac{\b}{L}}\, ,
\ee
where  $F_2\prt{\e_1,\e_2,\frac{\b}{L}}$ is an unknown function where its argument dimensions are normalized by the radius 
$R$ of the AdS space where needed, and $c_1,~c_2>0$ are constants that are specified numerically.
By adding higher powers of $L/\b$ to the fit it can be seen that the values of the coefficients $c_{1}, c_{2}$ are stable. We remark that
our results are valid in the low temperature limit, which is translated to $L\ll\b$. Moreover the fitted expression should be taken with caution due to the 
peculiarities of the numerics we have mentioned and the presence of the undetermined function $F$. We hope to return to this subject in
the future with more advanced results.

%%%%%%%%%%%%%%%%%
\section{Conclusions}
%%%%%%%%%%%%%%%%%

In this paper we have studied the properties of the flux tube connecting static quarks and anti-quarks in several cases. We have used two alternative approaches.

Firstly, working with generic confining theories we have calculated the flux generated by a heavy quark pair through
a probe Wilson loop. This is achieved by placing the measuring small Wilson loop in the middle of the heavy bound state
and at a distance above it. The resulting computation in the large $N$ strong coupling limit turns out to be a connected
minimal surface with boundaries the two Wilson loops.
The width of the tube in the middle of the pair can be found by specifying its minimal area. Using the properties of the
string solutions and the minimal surfaces in confining backgrounds, we have argued that the logarithmic broadening of the flux tube is a property of all confining theories in the context of the gauge/gravity duality. Our analysis is supported by numerical computations, where the system of equations is solved without any approximations.
Notice that our analysis is not applicable to conformal backgrounds and it will be surprising to us if the same properties
will be observed in backgrounds where confinement is not present. The reason is that the minimal surfaces will have different shape when confinement is not present.

In the large quark separation limit the rectangular Wilson loops can be
replaced by circular ones and this is not expected to affect the type of broadening. If one would like to repeat our analysis in non-confining 
theories with the presence of black hole at finite temperature, this assumption needs to checked.
It would be also challenging to build consistently the tubes in the Lifshitz-like theories and in
generic anisotropic theories at finite temperature, like for example in \cite{Mateos:2011ix}, where the calculation of several observables
involving the interaction between the heavy quark pair has been made \cite{Giataganas:2012zy,Giataganas:2013lga,Chernicoff:2012bu}.
Those backgrounds may be also considered as modified hard wall models to secure confinement. The anisotropy of space
will impose a non-trivial shape on the minimal connecting surface and it would be interesting to see to what extend the
logarithmic broadening of the width will hold. Notice that in the anisotropic theories have been found
several universal properties of the isotropic theories to be violated \cite{Rebhan:2011vd,Giataganas:2013hwa}.

As an alternative approach to understand the dynamics of confining strings we have considered fluctuations of strings in particular 
limits and in curved backgrounds. The strings have fixed end-points, and their time evolution leads to fluctuations in space time. 
We started with the time-independent fluctuations of strings with fixed end-points on the boundary. Then time-dependent fluctuations have been considered in curved backgrounds. 
The Green function of a differential equation with particular boundary conditions needs to be found in order to specify the width of the fluctuations. 
The setback is that the differential equations of the fluctuations can not be solved analytically. We have employed a point-splitting 
regularization of the Green function and carried out numerically the computation in the simplest case of the curved $AdS$ background. 
Although the dual theory is non-confining we computed an analogous quantity to the width of the heavy meson. There are difficulties in 
finding the two-point function even numerically, mostly due to the singularities which generate certain divergencies that need to be isolated. 
In the numerical fitting we were able to see that for small distances the quantity we have defined as width seems to have at least a logarithmically reducing term with the distance of the string, and that would be interesting to see if this is a generic property of non-confining theories. Due to some peculiarities of the numerics we plan to return to this result in the future for a more accurate treatment. As a next step one would like to do the full analysis of fluctuations in order to compute the width of flux tubes in confining theories.

An extension of our work would be to study the broadening of the tube between more involved heavy bound states formed 
in gauge/gravity duality than the quarkonia.  The strongly coupled bound state formed by k interacting heavy Q\={Q}, 
called k-strings is expected to have the logarithmic broadening for any confining theory. More demanding 
may be the holographic computation of the baryon effective width, which in effective string models using the 
$Y$ ansatz has been found to grow logarithmically \cite{Pfeuffer:2008mz}. Holographically the calculation maybe done with the
baryon vertex string \cite{Brandhuber:1998xy,Imamura:1998hf}, which already has ben found to have certain universal features among the different theories \cite{Giataganas:2012vw}. Moreover, it would be very interesting to relate the entropy grow associated with the heavy quark-antiquark bound state with the properties of the flux tube along the lines of \cite{Kharzeev:2014pha,Hashimoto:2014fha,Satz:2015jsa}.

{\bf Acknowledgements:} We are thankful to J. Casalderrey-Solana, V. Keranen, D. Mateos and M. Teper, for useful conversations and comments. The research of D.G is implemented under the "ARISTEIA" action of the "operational programme education and lifelong learning" and is co-funded by the European Social Fund (ESF) and National Resources. The research of N.I. is implemented under the ARISTEIA II action of the operational programme education and long life learning and is co-funded by the European Union (European Social Fund) and National Resources.

%%%%%%%%%%%%%%%%%%%%%%%%%%%%%%%%%%%%%%
\appendix{\section{An analytically solvable toy model}\label{app:QQtoy}
%%%%%%%%%%%%%%%%%%%%%%%%%%%%%%%%%%%%%%

In this section we use the mathematical formalism developed in the main text, to a toy model using a space-time string that leads to analytic results. 
The motivation is just to present an example where the Green function problem can be solved and its point splitting regularization can be carried out till the end analytically. 
So we present a toy model of a string embedded in the curved space, where the choice of its metric has been motivated by the fluctuation 
equations \eq{Dmain}, in order to obtain analytical solutions. The metric of the space is $ds^2= \sqrt{1+u} \prt{dx^2+dt^2}+(\sqrt{1+u})^{-1} du^2$. 
The analysis of the string is the same as in section \ref{stat:1} with the  difference that the $u$ direction has different scalings 
in the curved space where the string lives. In order to keep the length of the string finite in this case we need to introduce a large upper bound 
$M$ on $\s$ at the direction $u$, where we apply the boundary conditions. 
 In this simple space we find the fluctuations of the string. We point out that this is only a toy working model, which however allows to develop the 
 analytic methodology, and exhibits some interesting features of the defined ''width'', which we compute.
The turning point of the string is related to its length via eq. \eq{bbbc}, in this case
by 
\be
\s_0=\frac{L}{2 W_{-1}\prt{-L/4M}}\ , \la{s0w}
\ee 
where $W_{-1}\prt{x}$ is the generalized Lambert function.
The equation of the transverse fluctuations that need to be solved is \eq{unnor} in the form
\be
y'' +\frac{p'}{p} y' - {\Omega_n^2} \frac{r}{p} y = 0~, \label{Legendre2}
\ee
with
\bea
s = 2~,\quad r = \frac{x^2}{\sqrt{x^2-1}}~,\quad p &=& x^2 \sqrt{x^2-1}~.
\eea
We observe that the singularity at $\s=\s_0$, that is at $y=1$. The solution to eq. \eq{Legendre2} can be found if we set $y=(1-x^2)^{1/4}h\prt{x}/x$. Then we obtain for the unknown function $h$
\be
\frac{1}{x (1-x^2)^{3/4}} \left\{ (1-x^2)\, h'' -2x\, h' + \left[ \Omega_n^2 + \frac{3}{4} - \frac{1}{4(1-x^2)} \right]\, h
\right\} = 0\, .
\ee
which is the associated Legendre equation with solutions the associated Legendre functions
$P^{1/2}_{\l_n}$ and $Q^{1/2}_{\l_n}$, where $\l_n = -1/2 + K_n$ and $K_n =  \sqrt{1+\Omega_n^2}$. Therefore the differential equation eq. \eq{Legendre2} has the two independent sets of solutions
\bea
u_{1,n}(x) = \frac{(1-x^2)^{1/4}}{x} P_{\l_n}^{1/2},\hskip 0.5cm  u_{2,n}(x) = \frac{(1-x^2)^{1/4}}{x} Q_{\l_n}^{1/2}\, ,
\eea
which can be written in the simple algebraic form
\bea
P_{\l_n}^{1/2}(x) &=& \frac{1}{\sqrt{2\pi}} (x^2-1)^{-1/4}
\left\{  \left[ x+ \sqrt{x^2-1} \right]^{\l_n+1/2}  + \frac{1}{\left[ x+  \sqrt{x^2-1} \right]^{\l_n+1/2}}\right\}~,\nonumber\\
Q_{\l_n}^{1/2}(x) &=& i \sqrt{\frac{\pi}{2}} (x^2-1)^{-1/4} \frac{1}{\left[ x+  \sqrt{x^2-1} \right]^{\l_n+1/2}}\, .
\eea
By making the  change of variables $x = \cosh z~,$ with  $z\in [0,\infty)$
our variable $x$ is indeed $\in [1,\infty)$, we get the compact expressions
\bea
u_{1,n}(z)= \sqrt{\frac{2}{\pi}} \frac{\cosh(K_n z)}{\cosh z}~,\qquad
u_{2,n}(z)= \sqrt{\frac{\pi}{2}}\frac{e^{-K_n z}}{\cosh z}~.
\eea
The first set of solutions $u_{1,n}(z)$ is ruled out using the boundary conditions and requirement of normalizability and we end up to a Green function of the form
\bea
G(z,t; z',t') &=& \sum_n  \frac{ e^{-\Omega_n (t-t')/\s_0} + e^{-\Omega_n (\b - (t-t'))/\s_0} }{2\Omega_n\prt{1-e^{-\Omega_n \b/\s_0}}}\cdot\frac{K_n\,e^{-K_n z}e^{-K_n z'}}{\cosh z\cosh z'}~,
\eea
where we have used the generic function \eq{finalGreen} and $N_n^2=K_n$ by the normalization condition \eq{normf1}.

We continue with the point splitting regularization of $G$ by setting  $t' =  t + \e$, $z'  = z + \e/\s_0$
 and
adding and subtracting in the numerator a term in order to split $G$ into a sum of a $\b$-dependent and a $\b$-independent part. This leads to
\be\label{separation1}
w^2 = w^2_\infty + w_\b^2
\ee
with
\bea
w^2_\infty = \sum_n  \frac{K_n}{2\Omega_n}  e^{-(K_n \frac{\e}{\s_0} - \frac{\Omega_n}{\s_0})}+\cdots~, \qquad
w^2_\b = \sum_n  \frac{K_n}{\Omega_n} \frac{e^{-\Omega_n\frac{\b}{\s_0}- K_n\frac{\e}{\s_0} }\cosh (\Omega_n \frac{\e}{\s_0})}{1-e^{-\Omega_n \frac{\b}{\s_0}}} + \cdots~,\nonumber\\
\eea
where we have kept the leading terms from the expansion of the $\cosh \e$ dependent  terms.
In the limit of a continuous spectrum we take $\Omega_n \longrightarrow \omega$ and then to leading order we have
\bea
w^2_\infty = \int_0^\infty d\omega \frac{\sqrt{1+\omega^2}}{2 \omega} e^{- \frac{\e}{\s_0}[\sqrt{1+\omega^2}-\omega]}~,\quad
w^2_\b =  \int_0^\infty d\omega \frac{\sqrt{1+\omega^2}}{\omega} \frac{e^{-\frac{\b}{\s_0} \omega-\frac{\e}{\s_0}\sqrt{1+\omega^2}}}{1-e^{-\frac{\b}{\s_0} \omega}}~.\la{w12b}
\eea
We treat separately the two integrals, focusing first on $w^2_\infty$. With the change of variables $h=\sqrt{1+\o^2}-\o$ we obtain
\bea
w^2_\infty=\int_0^1 dh \frac{\prt{h^2+1}^2}{4 h^2\prt{h^2-1}}e^{-\frac{\e}{\s_0} h}\, ,\label{Ints}
\eea
where integrating the above and expanding in $\e$, we get elementary functions and the exponential integral function. In the small $\e$ limit  we arrive at
\be\la{wlog1}
w^2_\infty=\frac{1}{2}\prt{1-\log2-\lim_{h\rightarrow 0}\log h}-\frac{\e}{ \s_0}\prt{\frac{1}{8}+4 \log2+6 \lim_{h\rightarrow 0}\log h}\, .
\ee
Next we turn to $w^2_\b$, where we notice that the dominant contributions come from the $\o\to 0$ regime. 
By introducing a cut-off $r \rightarrow 0$ in the integral, the expression in \eq{w12b} can be written in the form
\bea
w^2_\b = \int_r^\infty d\omega \frac{\sqrt{1+\omega^2}}{\o}e^{-\frac{\e}{\s_0}\sqrt{1+\omega^2}} \sum_{n=0}^{\infty}  e^{- \frac{\b}{\s_0} \prt{n+1} \omega}~,
\eea
where the dominant contributions come from the $\o\to 0$ regime.
\begin{figure*}
\centerline{\includegraphics[width=80mm]{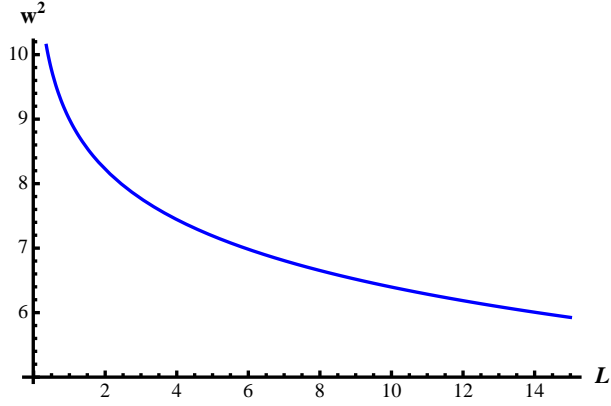}}
\caption{\small{The quantity $w^2$ in terms of $L$, where for small distances where our approximation is valid is reducing. 
For the plot we have fixed the normalized quantities $M=600$, $\b=400$, which gives $\s_0\ll M$.
}}\label{fig:w22}
\end{figure*}
We expand in $\e$ and we integrate  to obtain
\be
w^2_\b =\sum_ {n=1}^\infty\prt{\Gamma\left(0,\frac{\b n r}{\s_0}\right)+ \frac{\s_0^2}{2 \b^2 n^2}}\prt{1  -\frac{\e}{\s_0}}~,
\ee
that can be written in the more convenient form
\be
w^2_\b =\prt{\sum_ {n=1}^\infty E_1\left(\frac{\b n r}{\s_0}\right)+ \frac{\s_0^2}{2 \b^2}\zeta \prt{2}}\prt{1  -\frac{\e}{\s_0}}~,
\ee
where $\zeta(2)$ is the zeta function. Using the Euler-Maclaurin formula, we get for the leading contributions
\be
w^2_\b =\prt{\frac{\s_0}{\b r} + \g -1 + \ln \frac{\b}{\s_0} + \ln r+ \frac{\s_0^2}{2 \b^2}\zeta \prt{2}}\prt{1  -\frac{\e}{\s_0}}~,\label{wlog2}
\ee
with $\g$ the Euler constant. Isolating the divergences coming from the cut-offs we find from \eq{wlog1} \eq{wlog2} the total expression for the 'width'
\be\la{w222}
w^2=c_1-\ln \frac{\s_0}{\b}+\frac{\pi^2}{12} \frac{\s_0^2}{ \b^2}+\frac{\e}{\s_0}\prt{c_2+\ln \frac{\s_0}{\b}-\frac{\pi^2}{12} \frac{\s_0^2}{ \b^2}}~\, ,
\ee
with $c_1$ and $c_2$ constants that can be easily specified and $\s_0$ given by \eq{s0w}. 
In Figure \ref{fig:w22} we observe that at the small $L$ and $\s_0<M$ the quantity we treat as width is reduced in this case with the increase of length.

\bibliographystyle{JHEP}
\bibliography{botany}

\end{document}